# Entropy Transfer and Dynamics of Allostery in Proteins

Aysima Hacisuleyman and Burak Erman

*Department of Chemical and Biological Engineering, Koc University, Istanbul, Turkey*

Abstract

Allostery is an intrinsic spatiotemporal property of all proteins, resulting from long range correlations in the order of several nanometers and time scales of nanoseconds. Information is carried asymmetrically from one part to another by entropy transfer. Here, we present a master equation model of allosteric communication in proteins based on the transfer entropy concept of Schreiber (PRL, **85**, 465, 2000). We show how the model relates the path and velocity of asymmetric entropy transfer to conformational transitions over the rugged energy surface of proteins and how this relates to function.

Fluctuations of residues of a folded protein at equilibrium, experimentally identified as Debye-Waller factors, are manifestations of dynamic transitions between different conformational states of the protein. Few of these conformational transitions play significant role in protein function. Those that determine the allosteric communication patterns of the protein are of special interest. Signal transmission from one part of the protein to another, often to a distant point on the same protein, depends on the dynamics of conformations that include these points. Although allostery has been understood for more than sixty years [1] the fact that it is an intrinsic property of all proteins [2] and that it is a result of dynamic transitions over an ensemble of conformations is only recent [3-6]. NMR relaxation experiments now conclusively show that [7] allosteric communication proceeds through dynamic redistribution of the conformational ensemble. It is suggested that conformational changes are dominated by a series of local entropy fluctuations [8] playing the role of an entropic carrier of free energy [4]. Very recently, Carr–Purcell–Meiboom–Gill pulse sequence NMR measurements showed that entropy redistribution controls allostery in a protein [9]. Despite the strong experimental evidence that entropy transfer is the major process in allostery, there is no first principles statistical mechanical model that describes the relationship between entropy transfer and conformational transitions and the dynamics of allostery. The trajectories of two points that are active in allosteric communication are correlated and the uncertainty in one trajectory decreases due to the correlation, leading to transfer of entropy from one point to the other. Due to the transient nature of the correlations, entropy transfer is a function of time, i.e., the decrease of uncertainty in the present state of one trajectory results from previous values of the correlation with the other trajectory. In this letter, we present for the first time a Master Equation based model that uses molecular dynamics trajectories and characterizes entropy transfer in proteins. We employ the concept of entropy transfer introduced by Schreiber [10] and recently applied to proteins [11,12]. Following the historical practice [13] we use entropy transfer synonymously with information transfer or decrease in uncertainty. Therefore, when we say there is entropy transfer from trajectory $i$ to $j$, we mean that the uncertainty of $j$ is decreased due to the transfer from $i$. Entropy transfer calculated by the model is essentially the transfer on pre-existing equilibrium conformations [4] of proteins because the analysis is performed on equilibrium simulations on unperturbed proteins. Binding of a ligand, mutation of a residue, phosphorylation or any other perturbation at a specific point of a protein changes the dynamic conformational preferences. Those changes that affect the function of the protein through allosteric modulation are the significant ones because they directly affect human health. The present model explains the mechanism through which these effects are propagated and their consequences on protein behavior.

We apply the method to determine the most significant route of entropy transfer on a highly ubiquitous protein Ubiquitin and show that entropy transfer from one fluctuating quantity to another in the system exhibits strong causality which indicates directional filtering of information transfer in proteins.

The model

The model is based on the knowledge of the trajectory of the protein from which fluctuations of any given conformational variable may be calculated. Angle between two directions, a local volume, fluctuations of distances between residue pairs or of individual residues are a few of possible variables that may also be characterized experimentally. Here, we adopt distance trajectories, $d_i(t)$, between $n$ residue pairs $(n \leq (N-1)N/2)$ of a protein of $N$ residues. The set of $d_i(t)$'s at any time $t$ will constitute a state of the protein. The protein will be transforming from one set to another throughout the trajectory. The choice of $d_i(t)$ is a suitable indicator of interaction topologies and the ones that participate in information transfer and contribute to allosteric activity.

*Time dependent probabilities of $d_i(t)$*: The probability distribution of $d_i(t)$'s in general are single peaked and approximately normal. However, distances between certain pairs of residues have trajectories with multiple peaks, i.e., the pairs fluctuate around a mean distance and then there is a jump to a different distance and the fluctuations continue around the new mean distance. The peak values of $d_i(t)$ distributions may be regarded as isomeric states, the preference of one over the other affecting the function of the protein. Although a distance trajectory may possibly have several such isomeric states, usually there are two isomeric states, one where the two residues are close to and the other where they are far from each other. In general, only a small number, n, of distance pairs will exhibit isomerism that control the allosteric movement of the protein. We assume that n such distance trajectories exist, each having two isomeric states. The joint probability of observing n pairs in one of the two isomeric states is given by the probability density $p(d)=p(d_1,d_2,...,d_n)$. Each $d_i$ has two elements. In a recent work. LeVine and Weinstein named these two state system as the Ising model of allostery [6]. The dynamics of the transitions is obtained in full generality by the solution of the master equation, $\frac{\partial p(d)}{\partial t} = Ap(d)$. Here, the joint probabilities $p(d)$ are functions of time and $A_{kl}$ is the transition rate matrix from state l to k. Here, the state is the collection of the $n$ conformations, a total of $2^n$ in number. The solution of the master equation is [14,15]

$$p(d(t)) = p(d(t)|d(0))p(d(0)) \tag{1}$$

where, $p(d(t)|d(0))$ is the conditional probability of having $d_1(t), d_2(t), ... d_n(t)$ at time $t$, given that they were $d_1(0), d_2(0), ... d_n(0)$ at time 0. $p(d(t)|d(0))$ is expressed in terms of the eigenspectrum of $A$ as

$$p(d(t)|d(0)) = B \exp(Lt) B^{-1} \tag{2}$$

where, $B$ is the matrix of eigenvalues of $A$ and $L$ is the diagonal matrix of eigenvalues of $B$. Knowledge of the conditional probabilities allows for the evaluation of entropy transfer in the system [10]. Equation 2 gives the most general solution to the problem. Below, we will focus on two events, the distance trajectory for the ith pair and the other for the $j^{th}$ pair.

*Entropy transfer from one trajectory to another*: We are interested in evaluating the changes that will be induced in $d_i(t)$ by the presence of $d_j(t)$ and vice versa. This requires the knowledge of the amount of information transferred from one to the other throughout the trajectory and is measured using Shannon's entropy. According to Schreiber's work, transfer entropy from the ith pair to the $j^{th}$ pair with a time delay of $\tau$ can be written as

$$T_{i \to j}(t) = S(d_j(t)|d_j(0)) - S(d_j(t)|d_i(0), d_j(0)) \tag{3}$$

Here, $S(d_j(t)|d_i(0),d_j(0))$ is the conditional entropy of having state $d_j(t)$ at time $t$, given that the joint states of $i$ and $j$ are $d_i(0)$ and $d_j(0)$, respectively at time $0$. $S(d_j(t)|d_j(0))$ is the conditional entropy of having state $d_j(t)$ at time $t$, given that the state is $d_j(0)$ at time $0$. If states $i$ and $j$ are uncorrelated the two terms in Eq. 3 will be equal and entropy transfer will be zero. If correlated, then the earlier knowledge of state $d_i(t)$ will reduce the entropy $S(d_j(t)|d_i(0),d_j(0))$ below $S(d_j(t)|d_j(0))$, which means that the uncertainty in the trajectory of the $j^{th}$ pair will be reduced. Thus $T_{i \to j}(t)$ is a measure of the reduction of uncertainty in the fluctuations of state $j$ due to past values of $i$.

In the remaining part of the formulation, we write the variable $d_j(t)$ in terms of the discrete values $d_{jq}(t)$ where $q$ is the index referring to the isomeric state. Equation 3 then takes the form

$$T_{i \to j}(t) = -\sum_{q'} p(d_{jq'},0) \sum_{q} p(d_{jq},t|d_{jq'},0) \ln p(d_{jq},t|d_{jq'},0) + \\ \sum_{p'q'} p(d_{iq'},0;d_{jp'},0) \sum_{q} p(d_{jq},t|d_{iq'},0;d_{jp'},0) \ln p(d_{jq},t|d_{iq'},0;d_{jp'},0) \quad (4)$$

where,

$$p(d_{jq},t|d_{ip'},0;d_{jq'},0) = \sum_{p} p(d_{ip},t;d_{jq},t|d_{ip'},0;d_{jq'},0) \quad (5)$$

$$p(d_{jq},t|d_{jq'},0) = \sum_{p'} p(d_{jq},t|d_{ip'},0;d_{jq'},0) p(d_{ip'},0,d_{jq'},0) / p(d_{jq'},0) \quad (6)$$

In all calculations throughout the paper, entropy is taken in $k_B$ units. Equation 4 is the major result of the paper, using which we will determine entropy transfer in a protein and relate it to function.

*Example: Entropy transfer and function of Ubiquitin*: Ubiquitin (Ubq) is a globular protein, with the 3-dimensioanal Protein Data Structure (1UBQ.pdb) shown on the left panel of Figure 1. The last four C-terminal residues extend from the compact structure to form a tail, important for its function. The different conformations that the tail takes in interacting with target proteins and its position relative to specific residues of the protein itself determine the function of the protein [16,17]. We generated a 1 μs molecular dynamics trajectory of Ubq in water at 300 K (see details in Supplementary Material, SM). Atomic positions in the trajectory were aligned to the first frame, by using VMD to eliminate all translational and rotational degrees of freedom. Atomic positions and velocities were calculated by NAMD 2.11 at intervals of 2 fs and recorded in corresponding simulation output files with 2 ps intervals. Aligned Cartesian coordinates of alpha carbons were used to conduct modal analysis, and this analysis showed that the largest correlations in the system were between the C-terminal residue 76 and residues 9 and 53. Further, distances between the highly correlated residues were calculated by using the distance formula, $d_i = \sqrt{(x_1(t)-x_2(t))^2 + (y_1(t)-y_2(t))^2 + (z_1(t)-z_2(t))^2}$ where the subscripts 1 and 2 identify the two atoms at the extremities of the distance vector. The position of the two distances, 9-76 and 53-76 are shown in Figure 1, left panel. The probabilities are obtained by counting the number of occurrences in rectangular grids of size 0.72 for the 53-75 distance and 1.20 for the 9-76 distance. These grid values were obtained by dividing the total range in each direction into 20 grids. The probability distribution of the distance between 9 and 76 and 53 and 76 showed two peaks. Another pair, residues 39-75 also showed a non-Gaussian distribution of distance fluctuations, but the departure from Gaussian was weak and therefore is not taken as a variable of the problem. The probability surfaces were converted to energy surfaces according to $-\langle \ln(w(d_i,d_j)) \rangle$ (See SM). The energy surface for the joint distribution of the two distances is presented in Fig 1.

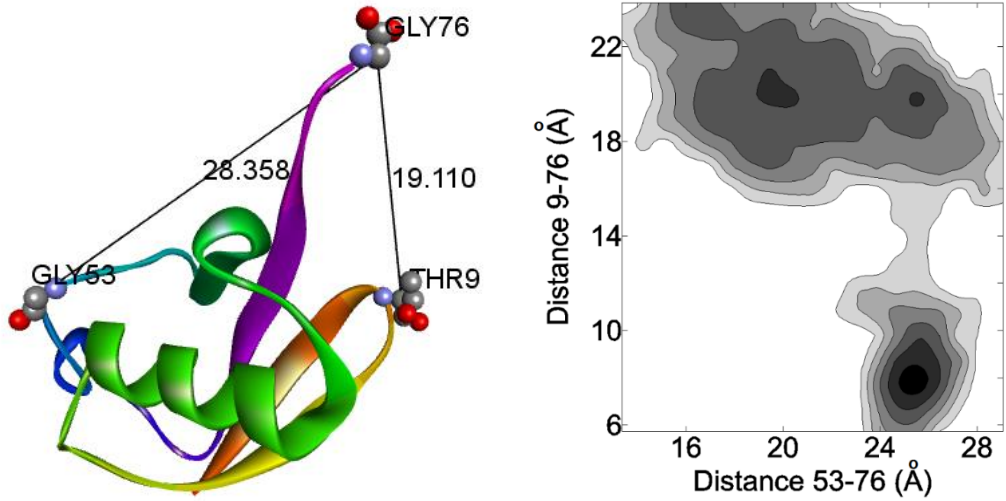

FIG 1. (Left panel) The 3-d structure of Ubiquitin. Crystal structure distances between 53-76 and 9-76 are indicated. (Right panel) Energy surface for the joint distribution of the distances 9-76 and 53-76. Darkest regions indicate minima.

We denote the pair 53-76 as $j$ and the pair 9-76 as $i$. $j$ has two states, corresponding to a larger $d_{j1}$ and smaller $d_{j2}$. Similarly, $i$ has a larger, $d_{i1}$, and close, $d_{i2}$, states. Thus, there are four combined states: $d_{i1}d_{j1}$ upper right minimum, $d_{i1}d_{j2}$ upper left minimum, $d_{i2}d_{j1}$ lower right minimum, $d_{i2}d_{j2}$ lower left minimum. The first three minima are seen in Fig 1, right panel. There is no minimum for the state $d_{i2}d_{j2}$.

The transition rate matrix $A$ was evaluated by using the Eyring equation [14], $r = r_0 \exp(-E_a/RT)$ (See SM). The saddle point energies $E_a$ between minima were obtained from Fig 1. The eigenvalues and eigenvectors of $A$ are used in Eq. 2 to obtain the conditional probability matrix. Together with the joint and marginal probabilities at time $0$ (See SM for numerical values), the conditional probabilities were used in Eq. 4 to obtain the entropy transfer between the two distances as shown in Fig 2.

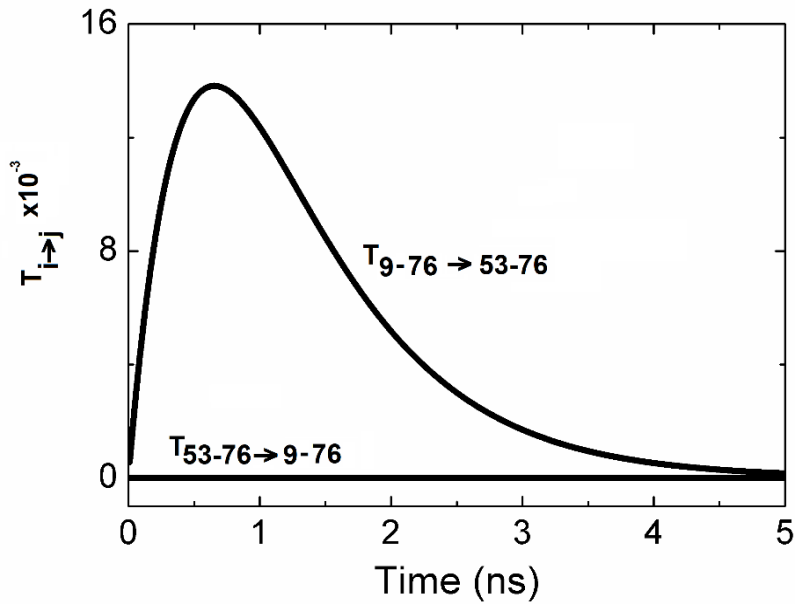

FIG 2. Entropy transfer between two important residue pairs of the protein.

According to Fig 2, fluctuations of the distance 9-76 decrease the uncertainty of the fluctuations of the distance 53-76, i.e., information is transferred from 9-76 to 53-76. The figure shows that no information is transferred from 53-76 to 9-76. Following our earlier work [11,12], we say that 9-76 drives the fluctuations of 53-76. The fact that 53-76 has no effect on 9-76 is an indication of strong causality in the problem where the direction of information flow is asymmetric and filtered out totally in one direction. The observed causality is associated with the characteristic energy landscape of the protein that favors certain conformational transitions and prevents others. Causality has been the focus of research in different fields such as economics [18], complex ecological systems [19], interacting oscillators [20], sensory motor networks [21], cardiac systems [22], etc. Causality-function relations in proteins is new. Recently, we detected strong causality in correlations among residue fluctuations in K-Ras, a protein whose loss of function is responsible for various forms of cancer, and related this causality to the function of the protein [23]. The dynamic aspect of entropy transfer is clearly seen in Fig 2. It is interesting to note that entropy transfer due to static correlations is zero because of the form of Eq. 3 (See SM). The amount of entropy transfer peaks around 0.6 ns and decays from thereon. The decay is completed around 5 ns, therefore the events taking place in the trajectory of 9-76 before 5 ns has no role on the present values of 53-76. The time $\tau^*$ of the peak value and the distance $\ell$ over which information is carried have been associated with the velocity $\ell/\tau^*$ of information transport in spatiotemporal systems [10,24-26]. The entropy transfer ordinate values shown in Figure 2 are for a pair of alpha carbon atoms, and correspond to energy values in the order of $10^{-2} k_B T$. If contributions from each degree of freedom of all-atom residues instead of only alpha carbons are considered, this value will be in the order of $k_B T$ due to the additivity of entropy.

In conclusion, the entropy transfer model for proteins can show the dynamic nature of asymmetric allosteric communication in folded proteins, and provides a simple recipe for calculations. Quantification of entropy transfer and causality and relating these to structure and function is of utmost importance for problems concerning human health because it is often the destruction of communication patterns of proteins by mutation that leads to various forms of disease, particularly cancer. This paper also directs attention to the role of quantitative dynamics, such as velocity of information transport in proteins. The present letter should be viewed as a proof of concept paper which opens the way to detailed computations on functional dynamic allostery in proteins.

# Supplementary material

# Entropy Transfer and Dynamics of Allostery in Proteins

Aysima Hacisuleyman and Burak Erman

*Department of Chemical and Biological Engineering, Koc University, Istanbul, Turkey*

In this Supplementary Material section, we give detailed description of the application of the model to the protein Ubiquitin. With the detailed explanations given in this part, one may apply the model to any other protein.

**Entropy transfer from i to j:** The basic equation of the paper given by Eq. 4 is:

$$
\begin{aligned}
T_{i \to j}(t) &= S(d_j(t)|d_j(0)) - S(d_j(t)|d_i(0), d_j(0)) = \\
&\quad -\sum_{q'} p(d_{jq'}, 0) \sum_{q} p(d_{jq}, t|d_{jq'}, 0) \ln p(d_{jq}, t|d_{jq'}, 0) \\
&\quad +\sum_{p'q'} p(d_{ip'}, 0; d_{jq'}, 0) \sum_{q} p(d_{jq}, t|d_{ip'}, 0; d_{jq'}, 0) \ln p(d_{jq}, t|d_{ip'}, 0; d_{jq'}, 0)
\end{aligned}
\quad (S1)
$$

where, $p(d_{jq}, t|d_{ip'}, 0; d_{jq'}, 0)$ and $p(d_{jq}, t|d_{jq'}, 0)$ are defined by Eqs. 5 and 6 in the paper. Transfer from *j* to *i* is obtained from S1 requires the change of indices as follows:

$$
\begin{aligned}
T_{j \to i}(t) &= S(d_i(t)|d_i(0)) - S(d_i(t)|d_j(0), d_i(0)) = \\
&\quad -\sum_{q'} p(d_{iq'}, 0) \sum_{q} p(d_{iq}, t|d_{iq'}, 0) \ln p(d_{iq}, t|d_{iq'}, 0) \\
&\quad +\sum_{p'q'} p(d_{iq'}, 0; d_{jp'}, 0) \sum_{q} p(d_{iq}, t|d_{iq'}, 0; d_{jp'}, 0) \ln p(d_{iq}, t|d_{iq'}, 0; d_{jp'}, 0)
\end{aligned}
\quad (S2)
$$

where

$$p\left(d_{iq},t\middle|d_{iq'},0;d_{jp'},0\right)=\sum_{p}p\left(d_{iq},t;d_{jp},t\middle|d_{iq'},0;d_{jp'},0\right) \quad (S3)$$

$$p\left(d_{iq},t\middle|d_{iq'},0\right)=\sum_{p'}p\left(d_{iq},t\middle|d_{iq'},0;d_{jp'},0\right)p\left(d_{iq'},0,d_{jp'},0\right)/p\left(d_{iq'},0\right) \quad (S4)$$

The different steps for determining the numerical values of the probabilities defined in Eqs. S1-S4 are as follows:

***Molecular dynamics simulations***: Initial coordinates of Ubiquitin were retrieved from Protein Data Bank with corresponding PDB code as 1UBQ. All-atom Molecular Dynamics simulation was performed by using NAMD 2.11 version with CHARMM22 force field parameters for Proteins and Lipids. Protein was immersed in TIP3P water-box and counter ions were placed to neutralize the system. RATTLE algorithm was used with time step of 2 fs and periodic boundary conditions were applied in an isothermal-isobaric NPT ensemble with constant temperature of 300 K and constant pressure of 1 bar. Temperature and pressure were controlled by Langevin thermostat and barostat, respectively. 1–4 scaling was applied to van der Waals interactions with a cutoff of 12.0 Å. System energy was minimized for 50 ps at 300 K and further subjected to MD production run for 1 µs. Trajectory was aligned to the first frame by using VMD 1.9.2 to eliminate all rotational and translational degrees of freedom and the analysis was conducted with the aligned Cartesian coordinates. Amplitude of fluctuations, $R_i(t)$ was calculated for each atom from the resultant MD trajectory. $\Delta R_i(t)$ was obtained by subtracting the mean of $i^{th}$ atoms amplitude of fluctuations from its $R_i(t)$.

***Choice of the variables of the protein***: In all calculations, we used only the alpha carbon of each residue. We performed modal decomposition of the trajectory according to the relation

$\Delta r(t) = \left\langle \Delta R(t) \Delta R(t)^T \right\rangle^{-1/2} \Delta R(t) = diag\,\lambda^{-1/2} V^T \Delta R(t)$ where $\Delta r(t)$ are the modal coordinates, $\lambda$ and V are the eigenvalues and eigen-vectors of $\left\langle \Delta R(t) \Delta R(t)^T \right\rangle$. We then chose the component $\Delta r_1(t)$ of the trajectory that corresponds to the largest eigenvalue of the correlation matrix $\left\langle \Delta R(t) \Delta R(t)^T \right\rangle$.

This mode shows the dominant motions of the protein and is the fundamental mode that shows the essential dynamics of the protein [27]. We calculated the mean square fluctuations of each residue in the first mode. In Figure 1, the mean square fluctuations of residues in the first mode are presented in terms of residue index.

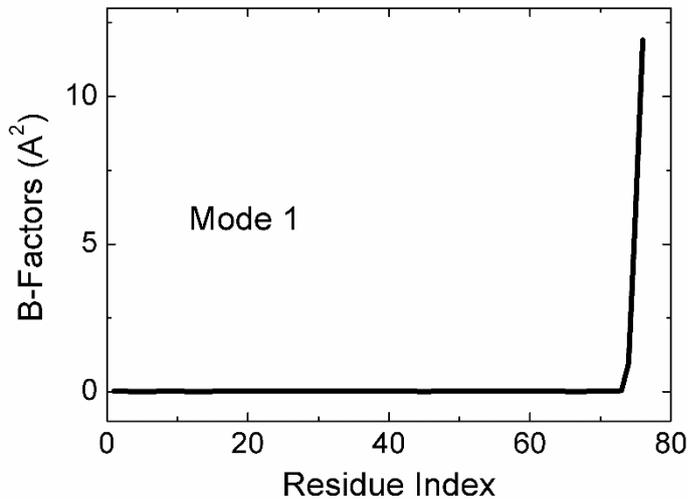

*Figure S1: B-factors of Ubq in the first mode.*

Figure S1 shows that only the C-terminal tail is active in the dominant mode. For this reason, we calculated the correlations in the first mode according to the relation $\langle \Delta r_1 \Delta r_1^T \rangle$. The dominant correlations are shown in figure S2:

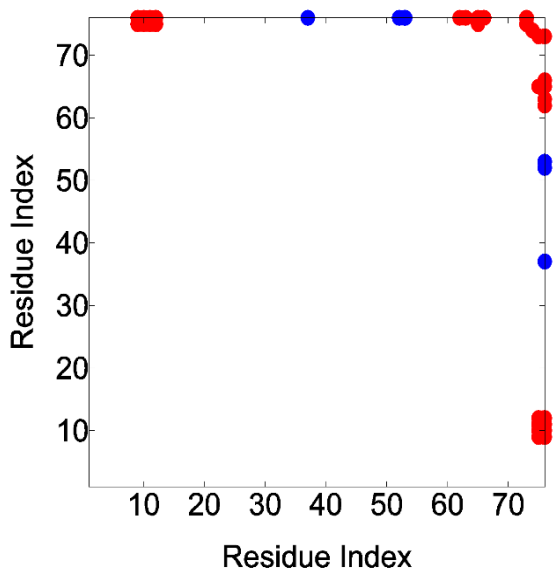

*Figure S2. Largest eigenvalue correlations of residue fluctuations.*

The largest negative correlation is obtained from Figure S2 is between residues 53 and 76. The largest positive correlation is between residues 9 and 76. We denote the distance between the pair 53-76 as $d_j$ and the distance between the pair 9-76 as $d_i$. These are the two dominant variables of the problem that are significant in controlling the dynamics of the protein. The distances are shown on the three dimensional structure of the protein, (PDB code 1UBQ.pdb) in Figure S3:

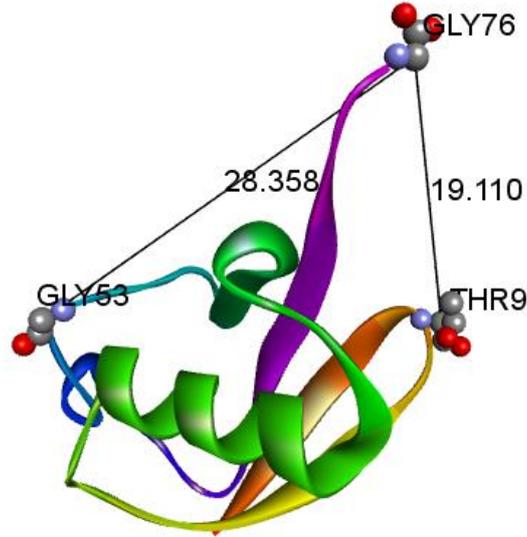

*Figure S3. The three dimensional structure of Ubiquitin (1UBQ.pdb) as ribbon diagram, showing the two distances between residues 9-76 and 53-76.*

**Calculation of the energy surfaces as the average log likelihood**: The energy surfaces are obtained according to the average log likelihood relation [28]. In one dimensional case, the relation is

$$-\langle \ln(w(d_k)) \rangle = -\int w(d_k) \ln w(d_k) \Delta(d_k) \approx -\frac{1}{N_T} \sum_1^{N_T} \ln(w(d_k)), \quad k=1,2 \quad \text{(S5)}$$

where, grids $\Delta(d_k)$ of size 0.72 for the 53-75 distance and 1.20 for the 9-76 distance are used.

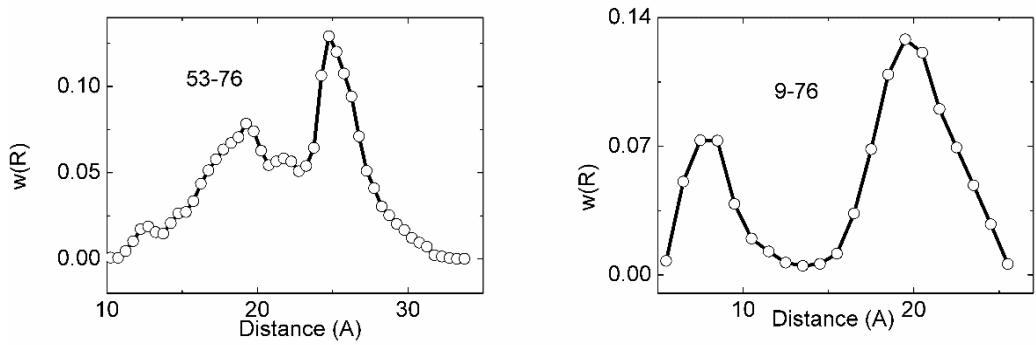

*Figure S4. Probability distribution of the distance 53-76, left panel, and 9-76, right panel.*

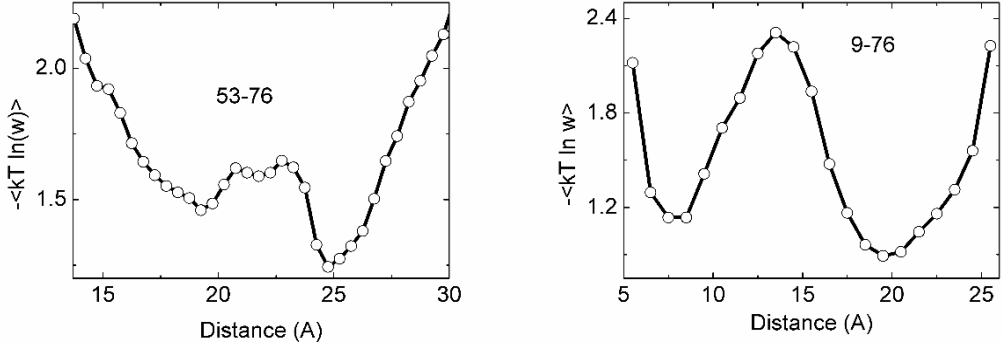

*Figure S5. Energies of distances, 53-76 left panel, 9-76 right panel.*

**Joint distributions:** The joint distribution of the distances 53-76 and 9-76 are obtained similarly, the average log likelihood term is used as

$$-\langle \ln(w(d_i,d_j)) \rangle = -\int w(d_i,d_j) \ln w(d_i,d_j) \Delta(d_i d_j) \approx -\frac{1}{N_T}\sum_1^{N_T} \ln(w(d_i,d_j)) \qquad (S6)$$

where $\Delta(d_i d_j)$ is a small region centered around $d_i d_j$, rectangular grids of size 0.72 for the 53-75 distance and 1.20 for the 9-76 distance were used. These were obtained by dividing the total range in each direction into 20 grids. The last sum is performed on the full trajectory with $N_T$ snapshots [28]. The joint energy surface is shown in Figure S6:

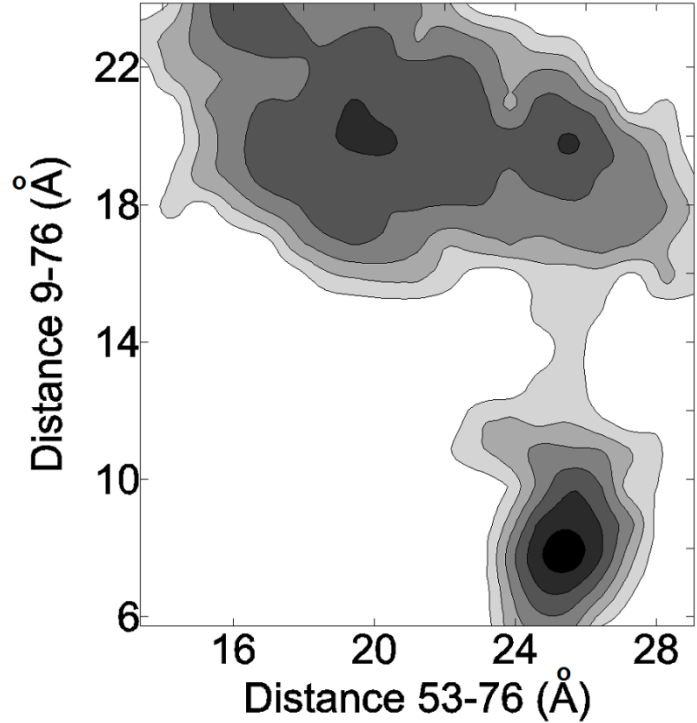

*Figure S6. Energy contours for the joint distribution of the distances 53-76 and 9-76.*

We let distance 53-75 be *j* and distance 10-76 be *i*. *j* has two states, a distant *j1* and closer *j2* state. Similarly, *i* has a distant, *i1*, and close, *i2*, states. Thus there are four combined states: *i1j1* upper right minimum, *i1j2* upper left minimum, *i2j1* lower right minimum, *i2j2* there is no minimum. Figure S7 shows the location of the states that are shown in Figure S6.

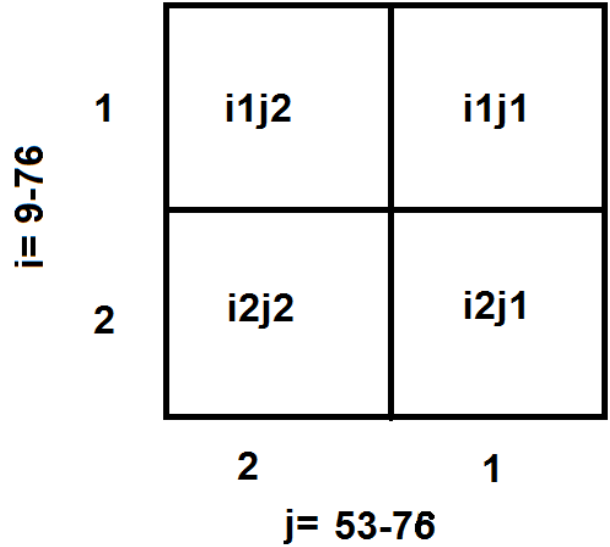

*Figure S7. Locations of the four minima for the two distances.*

The values of the joint probabilities are obtained as

$$p\begin{bmatrix} i1\,j1 \\ i1\,j2 \\ i2\,j1 \\ i2\,j2 \end{bmatrix} = \begin{bmatrix} 0.114 \\ 0.172 \\ 0.714 \\ 0 \end{bmatrix}$$

Marginal probabilities are obtained as $i1 = i1\,j1 + i1\,j2$, $i2 = i2\,j1 + i2\,j2$, $j1 = i1\,j1 + i2\,j1$ and $j2 = i1\,j2 + i2\,j2$.

$$p\begin{bmatrix} j1 \\ j2 \end{bmatrix} = \begin{bmatrix} 0.828 \\ 0.172 \end{bmatrix} \qquad p\begin{bmatrix} i1 \\ i2 \end{bmatrix} = \begin{bmatrix} 0.286 \\ 0.714 \end{bmatrix}$$

Comparison of the joint and marginal probabilities shows that the two distances follow dependent statistics.

***Transition rates:*** We calculate the transition rates using the Eyring equation [14]

$$r = (k_B T / h)(F^* \exp(-E_a / RT)) \tag{S7}$$

where h is Planck's constant, kB the Boltzmann constant, T absolute temperature, and $F^*$ is a measure of the curvature of the energy minima. $E_a$ is the energy to be overcome in passing from one minimum to the other. The energies are calculated from Figure S6 where saddle point passages from one minimum to the other are identified by recording several routes and the one with the lowest value of $E_a$ is chosen. The following are the energy maps obtained in this way:

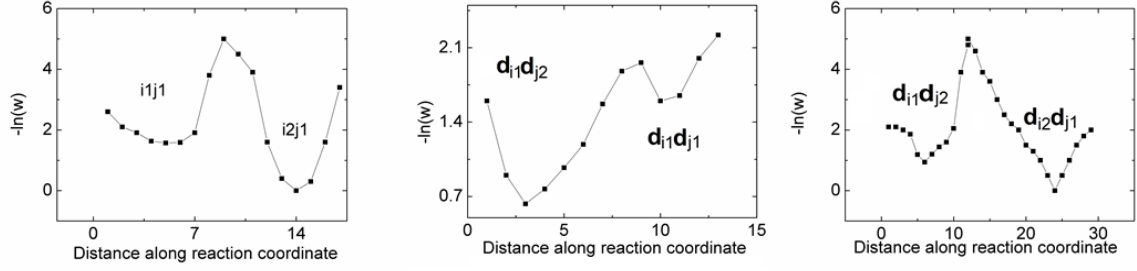

*Figure S8. Energy maps for the saddle point passages.*

The transition rate matrix, taking the front factor as 1 and RT-2.5 in Eq. S7 and using the energies shown in Figure S8 is obtained as:

$$A = \begin{array}{c} \phantom{11} \\ 11 \\ 12 \\ 21 \\ 22 \end{array} \begin{bmatrix} 11 & 12 & 21 & 22 \\ -1.010 & 0.873 & 0.137 & 0.000 \\ 0.859 & -0.996 & 0.137 & 0.000 \\ 0.254 & 0.197 & -0.451 & 0.000 \\ 0.000 & 0.000 & 0.000 & 0.000 \end{bmatrix}$$

The eigenvalues of the transition rate matrix are:

$$\lambda = \begin{bmatrix} -1.8690 \\ 0.0000 \\ -5.880 \\ 0 \end{bmatrix}$$

The components of the eigenvalue matrix V and V$^{-1}$ are

$$V = \begin{bmatrix} 0.7071 & -0.6508 & -0.3854 & 0 \\ -0.7071 & -0.6485 & -0.4307 & 0 \\ -0.0000 & -0.3947 & 0.8161 & 0 \\ 0 & 0 & 0 & 1.0000 \end{bmatrix}$$

$$V^{-1} = \begin{bmatrix} 0.7153 & -0.6989 & -0.0310 & 0 \\ -0.5903 & -0.5903 & -0.5903 & 0 \\ -0.2855 & -0.2855 & 0.9399 & 0 \\ 0 & 0 & 0 & 1.0000 \end{bmatrix}$$

*Vanishing of static entropy transfer*: We use the chain rule of conditional entropy

$$S(Y|X) = S(X,Y) - S(X)$$

and write the first line of Eq. S1 as

$$T_{i \to j}(t) = S(d_j(t), d_j(0)) - S(d_j(0)) - S(d_j(t), d_j(0), d_i(0)) + S(d_j(0), d_i(0))$$ (S5)

At t=0, the first two terms are equal, as well as the last two terms, leading to the vanishing of static entropy transfer